\documentclass[a4paper,11pt]{article}
\usepackage{pos}

\title{Solar Flare Hard X-ray Polarimetry with the CUbesat Solar Polarimeter (CUSP) mission}
\ShortTitle{Solar Flare Hard X-ray Polarimetry with CUSP}

\graphicspath{{Pictures}}

\author*[a]{Nicolas De Angelis}

\onbehalf{for the CUSP Team} 

\affiliation[a]{INAF-IAPS,\\
  100 via del Fosso del Cavaliere, Rome, Italy}

\emailAdd{nicolas.deangelis@inaf.it}

\abstract{The CUbesat Solar Polarimeter (CUSP) project is a CubeSat mission planned for a launch in low-Earth orbit and aimed to measure the linear polarization of solar flares in the hard X-ray band by means of a Compton scattering polarimeter. CUSP will allow us to study the magnetic reconnection and particle acceleration in the flaring magnetic structures of our star. CUSP is a project in the framework of the Alcor Program of the Italian Space Agency aimed at developing new CubeSat missions. It is undergoing a 12-month Phase B that started in December 2024.\\

The Compton polarimeter on board CUSP is composed of two acquisition chains based on plastic scintillators read out by Multi-Anode PhotoMultiplier Tubes for the scatterer part and GAGG crystals coupled to Avalanche PhotoDiodes for the absorbers. An event coincident between the two readout schemes will lead to a measurement of the incoming X-ray's azimuthal scattering angle, linked to the polarization of the solar flare in a statistical manner. The current status of the CUSP mission design, mission analysis, and payload scientific performance will be reported. The latter will be discussed based on preliminary laboratory results obtained in parallel with Geant4 simulations.}

\ConferenceLogo{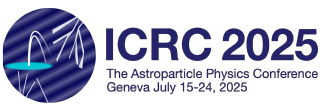}

\FullConference{39th International Cosmic Ray Conference (ICRC2025)\\
 15–24 July 2025\\
Geneva, Switzerland\\}

\begin{document}
\maketitle

\section{Motivation for Solar Flare Polarimetry}

Solar flares are intense bursts of electromagnetic radiation originating from the Sun's atmosphere, specifically in regions with strong magnetic fields, often near sunspots. These events are of crucial importance in solar physics, as they provide insight into the Sun's magnetic activity and the mechanisms governing energy release in the solar atmosphere.\\

Solar flares are fundamentally driven by the sudden release of magnetic energy stored in the Sun's corona, a process believed to be triggered by magnetic reconnection—where magnetic field lines rapidly realign and release energy. This phenomenon is central to solar physics, which studies the Sun's structure, dynamics, and magnetic behavior. Flares affect all layers of the solar atmosphere (photosphere, chromosphere, and corona), heating plasma to over $10^7$~K and accelerating particles to near-relativistic speeds.\\

While multi-wavelength observations have provided valuable information about these processes, many key questions remain unresolved. For instance, the precise geometry of the reconnection region, the efficiency of particle acceleration mechanisms, and the pitch-angle distribution of high-energy electrons are still debated. Spectroscopic and imaging data often leave ambiguities that prevent a full understanding of flare dynamics.\\

Polarimetry in the hard X-ray band offers a powerful and complementary tool to address these challenges. The degree and orientation of polarization are sensitive to the anisotropy of electron beams, the local magnetic field geometry, and the observer’s viewing angle \cite{Zharkova2010, Jeffrey2020}. Non-thermal bremsstrahlung, expected from energetic electrons during the impulsive phase of the flare \cite{Temmer2016, Nagasawa2022}, can reach polarization fractions of several tens of percent, while thermal bremsstrahlung is predicted to be only weakly polarized. Measurements of X-ray polarization can therefore disentangle thermal and non-thermal contributions, constrain the geometry of the acceleration region, and test theoretical predictions of electron beaming.\\

Despite its potential, solar flare polarimetry remains an under-explored frontier. Past measurements have been scarce and statistically limited, often providing only upper limits or tentative detections \cite{Tindo1970, Tindo1972a, Tindo1972b, Tramiel1984, Boggs2006, SuarezGarcia2006}. Dedicated instruments with improved sensitivity and temporal resolution are required to capture polarization signatures during the rapid evolution of solar flares. Progress in this area is not only crucial for fundamental heliophysics, but also has implications for space weather forecasting, as solar flares and associated coronal mass ejections are key drivers of disturbances in Earth’s near-space environment.\\

With the aim of performing spectro-polarimetric characterization of the hard X-ray emission from Solar Flares, we are developing a CubeSat mission called CUSP based on a dual-phase Compton polarimeter sensitive in the 25-100~keV band described hereafter.

\newpage
\section{The CUbesat Solar Polarimeter (CUSP)}
\subsection{The CUSP mission}

The CUbesat Solar Polarimeter (CUSP) is a CubeSat mission selected by the Italian Space Agency (ASI) in the frame of its Alcor program dedicated to nanosatellites. It consists of a 6U-XL platform developed by IMT s.r.l. hosting a single scientific payload. The main characteristics of the BUS can be appreciated in Table \ref{tab:platform_specs}. The payload is a dual phase Compton polarimeter that aims to perform Solar Flare polarimetry in the hard X-ray band, as described in the next section. The instrument is developed by the Italian National Institute for Astrophysics (INAF), while the flight electronics is designed by DEDA Connect s.r.l.. Moreover, the ground segment is located in Viterbo, Italy, and will be operated by the University of "La Tuscia".

\begin{table}[h!]
\centering
\begin{tabular}{|l|l|}
\hline
\textbf{Peak Power} & $\sim 30~\mathrm{W}$ with Deployable Panels in Sun Pointing \\ \hline
\textbf{Battery} & Up to 84 Wh \\ \hline
 & $< 0.04^\circ~@~3\sigma$ (AKE) \\
\textbf{Attitude accuracy} & $< 0.08^\circ~@~3\sigma$ (APE) \\
 & $< 2 ^\circ/s$ Slew Rate \\ \hline
\textbf{Operative frequencies} & S-Band downlink; UHF-Band uplink / downlink \\ \hline
\textbf{Downlink throughput} & Up to 5 (10) Mbps nominal (max) \\ \hline
\textbf{Available interfaces} & CAN Bus, I2C, UART, SPI, RS485 \\ \hline
\textbf{Regulated bus} & 3.3V, 5V \& 12V \\ \hline
\textbf{Not regulated bus} & 16V (12V–16.8V) \\ \hline
\textbf{Available volume for the payload} & 2.5U \\ \hline
\textbf{Nominal life time} & 3 years in LEO \\ \hline
\end{tabular}
\caption{Key technical specifications of the platform.}
\label{tab:platform_specs}
\end{table}

Although the current baseline for the CUSP mission is to have a single satellite, a goal has been set to potentially launch a pair of CubeSats to increase the Sun's coverage. This would allow to increase the time fraction that we are observing compare to the single satellite solution by a significant amount, namely from 45~\% to 68~\% for a Moon-Midnight (12:00 LTAN) or a Mid-Morning (10:00 LTAN) orbit, and from 69~\% to 88~\% for a Dawn-Dusk (06:00 LTAN) orbit.\\

The project started its Phase B in December 2024 and concluded its System Requirements Review in July 2025. It is now going towards the end of its Phase B, with the Preliminary Design Review currently planned for December 2025. The current timeline foresees a launch in the late 2027/early 2028 period. More details about the satellite BUS, ground segment, mission analysis, and operation concept can be found in \cite{CUSP_SPIE25}.

\newpage
\subsection{CUSP's Hard X-ray Polarimeter}

\begin{figure}[h!]
\centering
\includegraphics[height=.42\textwidth]{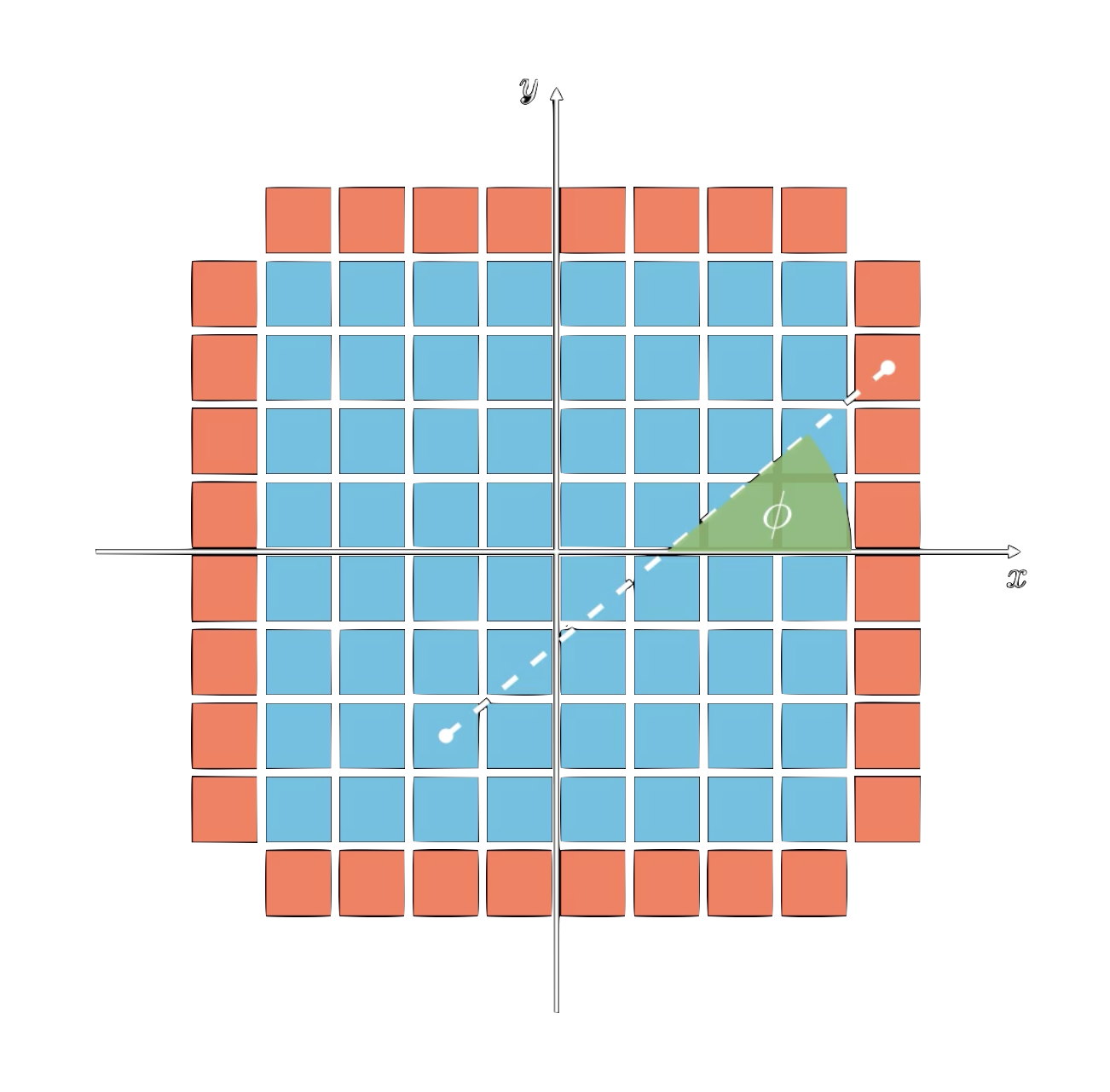}\includegraphics[height=.42\textwidth]{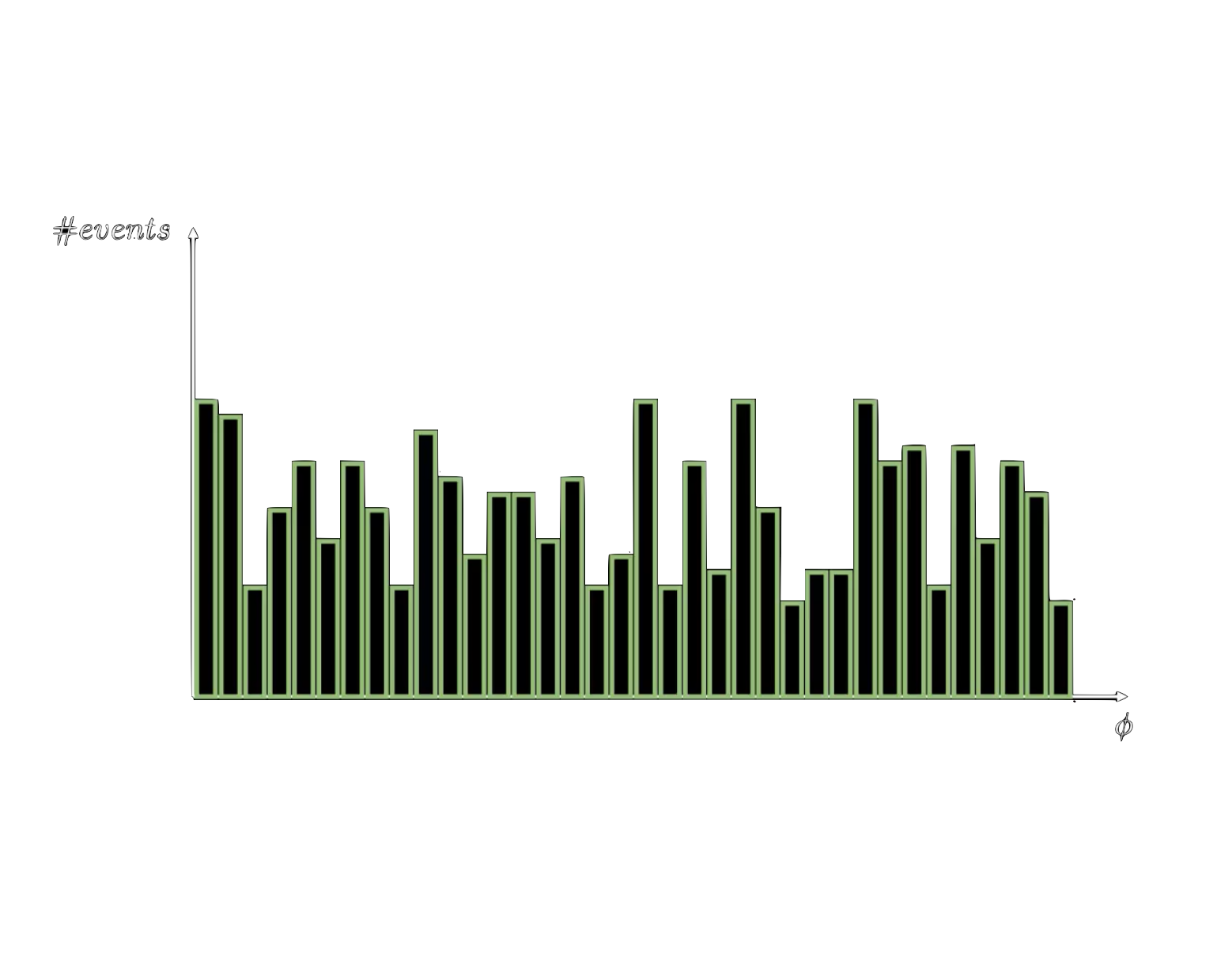}
 \caption{Detection principle of a dual-phase Compton polarimeter. \textbf{Left:} Top view of CUSP with 64 scatters in blue surrounded by 32 absorbers in red. The dashed white line shows an event for which the incoming photon is scattered and then absorbed in two different channels (adapted from \cite{CUSP_animation}, with permissions). The azimuthal scattering direction is given by the angle $\phi$. \textbf{Right:} Azimuthal scattering angle distribution, a.k.a. modulation curve, built from a non-polarized incoming flux (adapted from \cite{CUSP_animation}, with permissions).}
 \label{fig:compton_polarimetry}
\end{figure}

CUSP's payload is a dual-phase Compton polarimeter operating in the 25--100 keV band. Its principle is illustrated in Figure~\ref{fig:compton_polarimetry}. Incoming photons first scatter in a low-$Z$ plastic scintillator, then are absorbed in a high-$Z$ GAGG:Ce crystal. The azimuthal distribution of such coincidence events encodes the polarization, as linearly polarized photons scatter preferentially perpendicular to their polarization vector. This behavior is described by the Klein-Nishina cross section:

\begin{equation}
\frac{d\sigma}{d\Omega} = \frac{r_e^2}{2}\,\left(\frac{E'}{E}\right)^2
\left( \frac{E}{E'} + \frac{E'}{E} - 2 \sin^2\theta \cos^2\phi \right),
\end{equation}

where $E$ and $E'$ are the incident and scattered photon energies, $\theta$ is the polar scattering angle, and $\phi$ the azimuthal angle relative to the polarization vector.

The resulting histogram of $\phi$ angles is known as the \emph{modulation curve}. Its amplitude, normalized to 100\% polarized radiation, defines the modulation factor $\mu_{100}$, which together with efficiency $\epsilon$ gives the polarimeter quality factor $Q=\mu\sqrt{\epsilon}$. The relative amplitude of the measured modulation for a given source gives its polarization degree (PD), white its phase is linked to the polarization angle (PA).

\begin{figure}[h!]
\centering
\includegraphics[width=.9\textwidth]{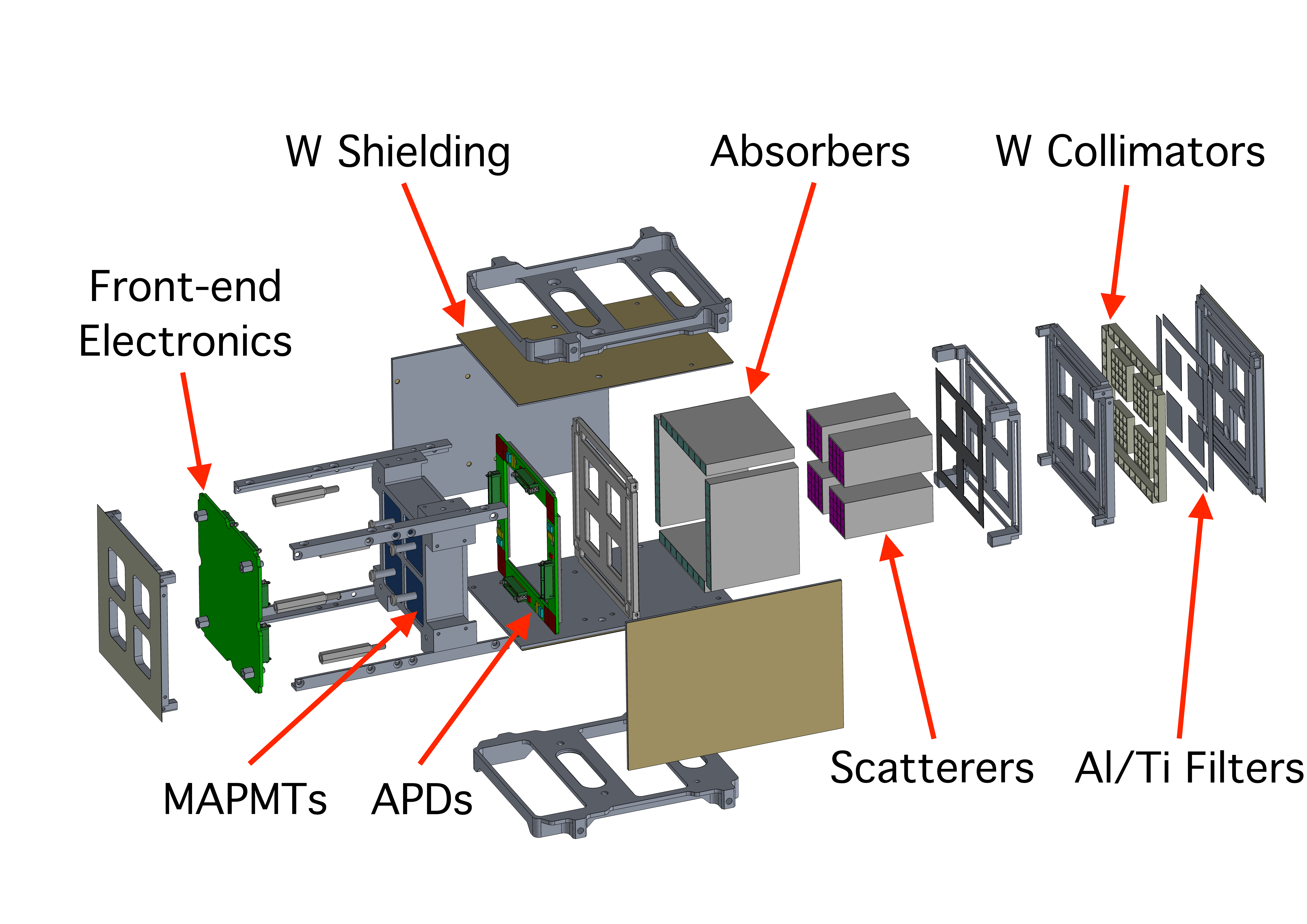}
 \caption{CAD exploded view of CUSP's hard X-ray polarimeter}
 \label{fig:CAD_CUSP}
\end{figure}

The polarimeter of CUSP comprises 64 plastic scintillator bars read out by 4 multi-anode photomultiplier tubes (MAPMTs), surrounded by 32 GAGG crystals read out by avalanche photodiodes (APDs). A tungsten collimator restricts the field of view to $\pm\,36^\circ$, while the zenith of the instrument is always pointing towards the Sun. The MAPMTs and APDs are read out by custom front-end electronics based on the MAROC-3A and SKIROC-2A ASICs from Weeroc. An exploded view of the CAD design of the payload is shown in Figure \ref{fig:CAD_CUSP}.

While functional and preliminary performance tests have been conducted using development boards coupled to single channel sensors and scintillators, as reported in  \cite{CUSP_SPIE25_proto, CUSP_ASAPP}, a more representative prototype is under development. It will be characterized and tested by the end of the project's Phase B and is currently under construction at INAF-IAPS. A structural model is also being developed to validate the mechanical design of the payload through vibration testing.

Simulations and laboratory prototypes show that CUSP will achieve Minimum Detectable Polarizations (MDP) of a few percent in minutes of integration. The MDP of a polarimeter is the minimum true polarization that a source should have in order to obtain a significant measurement of its polarization degree with a 99~\% confidence level \cite{Weisskopf2010, Strohmayer2013}. Table~\ref{tab:mdp} lists benchmark estimates of the MDP based on \cite{SaintHilaire2008}, while more details are provided in \cite{CUSP_SPIE25}.

\begin{table}[h]
\centering
\begin{tabular}{lccc}
\hline
Flare class & Integration time (s) & MDP (\%) \\
\hline
M5.2 & 284 & 7.8 \\
X1.2 & 240 & 3.9 \\
X10 & 351 & 0.9 \\
\hline
\end{tabular}
\caption{Estimated MDP at 99\% confidence level for different flare classes in the 25--100 keV band.}
\label{tab:mdp}
\end{table}

\newpage
\section{Conclusions and Outlook}

The CUbesat Solar Polarimeter, or CUSP, is a 6U-XL CubeSat mission currently in its Phase B that aims to measure the polarization of hard X-ray photons from solar flares in order to better understand magnetic reconnection and particle acceleration processes at play in the Sun's atmosphere. It will also play a crucial role in better understanding the link between solar flares and coronal mass ejection, thereby contributing to the field of space weather.\\

The Compton polarimeter consists of an 8$\times$8 array of plastic scintillators surrounded by 4 strips of 8 GAGG crystals. Thanks to its optimized design, it will reach unprecedented sensitivities to solar flares, allowing significant polarization measurements of the hard X-ray emissions from bright events.\\

A first representative prototype of the polarimeter is currently under construction for a better assessment of the instrument's scientific performances. The mission is currently targeting a launch in end-2027/early-2028, contingent on programmatic progress and institutional alignment.

\section*{Acknowledgments}

This work is funded by the Italian Space Agency (ASI) within the Alcor Program, as part of the development of the CUbesat Solar Polarimeter (CUSP) mission under ASI-INAF contract n. 2023-2-R.0.

\clearpage
\section*{Full Authors List: CUSP Team}

%
\scriptsize
\noindent
Nicolas De Angelis$^1$,
Andrea Alimenti$^{1,2}$,
Davide Albanesi$^3$,
Ilaria Baffo$^4$,
Daniele Brienza$^5$,
Riccardo Campana$^6$,
Valerio Campamaggiore$^3$,
Mauro Centrone$^7$,
Enrico Costa$^1$,
Giovanni Cucinella$^8$,
Andrea Curatolo$^9$,
Giovanni De Cesare$^6$,
Giulia de Iulis$^3$,
Ettore Del Monte$^1$,
Andrea Del Re$^3$,
Sergio Di Cosimo$^1$,
Simone Di Filippo$^8$,
Giuseppe Di Persio$^1$,
Immacolata Donnarumma$^5$,
Sergio Fabiani$^1$,
Pierluigi Fanelli$^4$,
Nicolas Gagliardi$^{10}$,
Abhay Kumar$^1$,
Alessandro Lacerenza$^1$,
Paolo Leonetti$^3$,
Pasqualino Loffredo$^1$,
Giovanni Lombardi$^{1,11}$,
Matteo Mergè$^5$,
Gabriele Minervini$^{12}$,
Dario Modenini$^{9,10}$,
Fabio Muleri$^1$,
Andrea Negri$^8$,
Daniele Pecorella$^9$,
Massimo Perelli$^8$,
Alice Ponti$^{10}$,
Paolo Romano$^{13}$,
Alda Rubini$^1$,
Emanuele Scalise$^1$,
Enrico Silva$^{1,2}$,
Paolo Soffitta$^1$,
Paolo Tortora$^{9,10}$,
Alessandro Turchi$^5$,
Valerio Vagelli$^5$,
Emanuele Zaccagnino$^5$,
Alessandro Zambardi$^3$,
and
Costantino Zazza$^4$\\

\noindent
$^1$INAF-IAPS, 100 via del Fosso del Cavaliere, Rome, Italy.\\
$^2$Department of Industrial, Electronic and Mechanical Engineering, "Roma Tre" University, via V. Volterra 62, 00146 Rome, Italy.\\
$^3$DEDA Connect s.r.l., Via Vincenzo Lamaro 51, 00173 Roma, Italy.\\
$^4$DEIM, Università degli studi della Tuscia, Largo dell’Università, 01100 Viterbo, Italy.\\
$^5$ASI, Via del Politecnico snc, 00133, Roma, Italy.\\
$^6$INAF-OAS Bologna, via Pierobetti 93/3, 40129 Bologna, Italy.\\
$^7$INAF-OAR, Via Frascati 33, 00040, Monte Porzio Catone, Italy.\\
$^8$IMT s.r.l., via Carlo Bartolomeo Piazza 30, 00161 Rome, Italy.\\
$^9$Department of Industrial Engineering - Alma Mater Studiorum Università di Bologna, Via Montaspro 97, 47121 Forlì, Italy.\\
$^{10}$Interdepartmental Centre for Industrial Aerospace Research - Alma Mater Studiorum Università di Bologna, Via Carnaccini 12, 47121 Forlì, Italy.\\
$^{11}$Dipartimento di Ingegneria dell’Impresa "Mario Lucenti", Università degli Studi di Roma Tor Vergata, Via Cracovia 50, 00133 Roma, Italy.\\
$^{12}$INAF-Headquarters, Viale del Parco Mellini 84, 00136, Roma, Italy.\\
$^{13}$INAF-OACT, Via S. Sofia 78, 95123, Catania, Italy.


\begin{thebibliography}{99}


\bibitem{Zharkova2010}

Zharkova, V. V., A. A. Kuznetsov, and T. V. Siversky. \textit{"Diagnostics of energetic electrons with anisotropic distributions in solar flares-I. Hard X-rays bremsstrahlung emission."} \href{https://doi.org/10.1051/0004-6361/200811486}{Astronomy \& Astrophysics 512 (2010): A8}.


\bibitem{Jeffrey2020}

Jeffrey, Natasha LS, Pascal Saint-Hilaire, and Eduard P. Kontar. \textit{"Probing solar flare accelerated electron distributions with prospective X-ray polarimetry missions."} \href{https://doi.org/10.1051/0004-6361/202038626}{Astronomy \& Astrophysics 642 (2020): A79}.

\bibitem{Temmer2016}

Temmer, Manuela. \textit{"Kinematical properties of coronal mass ejections."} \href{https://doi.org/10.1002/asna.201612425}{Astronomische Nachrichten 337.10 (2016): 1010-1015}.

\bibitem{Nagasawa2022}

Nagasawa, Shunsaku, et al. \textit{"Study of Time Evolution of Thermal and Nonthermal Emission from an M-class Solar Flare."} \href{https://doi.org/10.3847/1538-4357/ac7532}{The Astrophysical Journal 933.2 (2022): 173}.

\bibitem{Tindo1970}

Tindo, I. P., et al. \textit{"On the polarization of the emission of X-ray solar flares."} \href{https://doi.org/10.1007/BF00240179}{Solar Physics 14.1 (1970): 204-207}.

\bibitem{Tindo1972a}

Tindo, I. P., et al. \textit{"New measurements of the polarization of X-ray solar flares."} \href{https://doi.org/10.1007/BF00153385}{Solar Physics 24.2 (1972): 429-433}.

\bibitem{Tindo1972b}

Tindo, I. P., et al. \textit{"Preliminary interpretation of the polarization measurements performed on ‘Intercosmos-4’during three X-ray solar flares."} \href{https://adsabs.harvard.edu/full/1972SoPh...27..426T}{Solar Physics 27.2 (1972): 426-435}.

\bibitem{Tramiel1984}

Tramiel, Leonard J., R. Novick, and G. A. Chanan. \textit{"Polarization evidence for the isotropy of electrons responsible for the production of 5-20 keV X-rays in solar flares."} \href{https://adsabs.harvard.edu/full/1984ApJ...280..440T}{Astrophysical Journal, Part 1 (ISSN 0004-637X), vol. 280, May 1, 1984, p. 440-447. 280 (1984): 440-447}.

\bibitem{Boggs2006}

Boggs, Steven E., W. Coburn, and E. Kalemci. \textit{"Gamma-ray polarimetry of two X-class solar flares."} \href{https://doi.org/10.1086/498930}{The Astrophysical Journal 638.2 (2006): 1129}.

\bibitem{SuarezGarcia2006}

Suarez-Garcia, Estela, et al. \textit{"X-ray polarization of solar flares measured with RHESSI."} \href{https://doi.org/10.1007/s11207-006-0268-1}{Solar Physics 239.1 (2006): 149-172}.

\bibitem{CUSP_SPIE25}

Fabiani, Sergio, et al. \textit{"The CUbesat Solar Polarimeter (CUSP): mission overview II."} \href{https://arxiv.org/abs/2508.00661}{SPIE Optics and Photonics, 2025, Accepted}.

\bibitem{CUSP_animation}

De Angelis, Nicolas \textit{"Dual phase Compton polarimeter working principle."} \href{https://doi.org/10.5281/zenodo.17017074}{Zenodo, 2025}.

\bibitem{CUSP_SPIE25_proto}

De Angelis, Nicolas, et al. \textit{"Prototype Development and Calibration of the CUbesat Solar Polarimeter (CUSP)."} \href{https://arxiv.org/abs/2508.00642}{SPIE Optics and Photonics, 2025, Accepted}.

\bibitem{CUSP_ASAPP}
De Angelis, Nicolas, et al. \textit{"Spectral performances of single channel plastic and GAGG of the CUbesat Solar Polarimeter (CUSP) for Heliophysics and Space Weather."} In preparation, 2025.

\bibitem{Weisskopf2010}
Weisskopf, Martin C., Ronald F. Elsner, and Stephen L. O'Dell. \textit{"On understanding the figures of merit for detection and measurement of x-ray polarization."} \href{https://doi.org/10.1117/12.857357}{Space Telescopes and Instrumentation 2010: Ultraviolet to Gamma Ray. Vol. 7732. SPIE, 2010}.

\bibitem{Strohmayer2013}
Strohmayer, Tod E., and Tim R. Kallman. \textit{"On the statistical analysis of X-ray polarization measurements."} \href{https://www.doi.org/10.1088/0004-637X/773/2/103}{The Astrophysical Journal 773.2 (2013): 103}.

\bibitem{SaintHilaire2008}
Saint-Hilaire, Pascal, Sam Krucker, and Robert P. Lin. \textit{"A statistical survey of hard X-ray spectral characteristics of solar flares with two footpoints."} \href{https://doi.org/10.1007/s11207-008-9193-9}{Solar Physics 250.1 (2008): 53-73}.


\end{thebibliography}
\end{document}